\documentclass[preprint]{elsarticle}

\newcommand{\Rvec}{{\bf R}}

\newcommand{\evec}{{\bf e}}
\newcommand{\cvec}{{\bf c}}

\newfont{\bg}{cmr10 scaled\magstep4}
\newcommand{\bigzerol}{\smash{\hbox{\bg 0}}}
\newcommand{\bigzerou}{\smash{\lower1.7ex\hbox{\bg 0}}}

\begin{document}

\title{Iterative diagonalization of symmetric matrices in mixed precision}

\author{Eiji Tsuchida\corref{COR1}}
\ead{Eiji.Tsuchida@aist.go.jp}
\cortext[COR1]{Corresponding author}

\author{Yoong-Kee Choe}

\address{Nanosystem Research Institute, 
   National Institute of Advanced Industrial Science and Technology (AIST), 
   Tsukuba Central 2, Umezono 1-1-1, Tsukuba 305-8568, Japan}

\begin{abstract}
Diagonalization of a large matrix is the computational bottleneck 
in many applications such as electronic structure calculations. 
We show that a speedup of over 30\% can be achieved 
by exploiting 32-bit floating point operations, while keeping 64-bit accuracy. 
Moreover, most of the computationally expensive operations are 
performed by level-3 BLAS/LAPACK routines in our implementation, 
thus leading to optimal performance on most platforms. 
Further improvement can be made by using problem-specific preconditioners 
which take into account nondiagonal elements. 
\end{abstract}


\maketitle

\section{Introduction}
Matrix diagonalization plays an important role in many areas of science and engineering. 
In electronic structure calculations of complex systems, for instance, 
most of the computational effort is spent on 
the numerical solution of the Schr\"odinger equation at various levels of approximation. 
This procedure is equivalent to an eigenvalue problem in which 
only a small subset of the eigenvalues 
of a large symmetric matrix is desired \cite{DFTREV1,DFTREV2,DFTREV3}. 
To this end, iterative diagonalization is more favorable than direct methods \cite{TMPL} 
which are designed for calculating all eigenvalues of dense matrices. 

Numerical calculations on modern computers are generally performed using 64-bit 
double precision (DP) floating-point numbers, which are accurate to 15-16 significant digits. 
On the other hand, 32-bit single precision (SP) floating-point numbers 
with 7-8 digits of accuracy are more efficient in terms of computational cost, 
memory usage, network bandwidth, and disk storage \cite{LIN2}. 
In recent years, considerable effort has been made 
to obtain the results with DP accuracy at the expense of SP operations. 
In particular, a variety of {\it mixed precision} algorithms 
have been developed for the solution of linear equations \cite{LIN2,LIN1}. 

Similarly, several researchers have attempted to solve eigenvalue problems 
with mixed precision algorithms in the past \cite{DIAG1,DIAG2}. 
Unfortunately, most of these approaches correspond to 
direct methods for dense matrices, and thus are of limited use 
in electronic structure calculations. 
In this paper, we show how to exploit the mixed precision arithmetic 
for iterative solution of large-scale eigenvalue problems, 
with special emphasis on electronic structure calculations.

\section{Theory}
\subsection{Trace minimization method}
\label{TMMSEC}
Let ${\cal H}$ be a real, symmetric matrix of dimension $N$. 
The eigenvalues and eigenvectors of ${\cal H}$ are defined by 
\begin{equation}
\label{BASICEQ}
{\cal H} \evec_j = \lambda_j \evec_j, \quad \quad j=1,2,\cdots,N, 
\end{equation}
where $\lambda_1 \le \lambda_2 \le ... \le \lambda_{N}$, and  
$\evec_1, \evec_2, \cdots, \evec_N$ form a set of orthonormal vectors. 
We also assume that ${\cal H}$ is a sparse matrix with $O(N)$ nonzero entries. 

Our aim is to calculate the $m$ lowest eigenvalues of ${\cal H}$ and 
the corresponding eigenvectors, where $1 \ll m \ll N$, and typically, 
$m= 10^2$-$10^3$ and $N = 10^5$-$10^6$ in electronic structure calculations \cite{NAFION,SPES}. 
To be precise, it is often sufficient to calculate only the sum of the eigenvalues, 
\begin{equation}
\label{EGEQ}
E_{\rm G} = \sum_{j=1}^m \lambda_j,  
\end{equation}
and the subspace spanned by $\evec_1, \evec_2, \cdots, \evec_m$, 
where $E_{\rm G}$ corresponds to the ground-state energy of the system \cite{DFTREV1,DFTREV2,DFTREV3}. 
Hereafter we assume the presence of a gap in the spectrum, i.e., 
\begin{equation}
\epsilon_{\rm gap} = \lambda_{m+1} -\lambda_m > 0, 
\end{equation}
which is satisfied in nonmetallic systems \cite{DFTREV1,DFTREV2,DFTREV3}. 

The trace minimization method \cite{DFTREV3,TRMIN} is based on the fact that if 
\begin{equation}
\label{TRHEQ}
E (\cvec_1, \cvec_2, \cdots , \cvec_m) = \sum_{j=1}^m \cvec_j^T {\cal H} \cvec_j
\end{equation}
is minimized subject to the orthonormality conditions 
\begin{equation}
\label{CIJEQ}
\cvec_i^T \cvec_j = \delta_{ij}, \quad \quad i,j=1,2,\cdots,m, 
\end{equation}
$E = E_{\rm G}$ holds at the minimum, and 
$ C = \left( \cvec_1 \,\, \cvec_2 \,\, \cdots \,\, \cvec_m \right)$ 
spans the same subspace as $\left\{ \evec_1, \evec_2, \cdots, \evec_m \right\}$. 
In matrix form, Eqs. (\ref{TRHEQ}) and (\ref{CIJEQ}) can be written as 
\begin{equation}
E (C) = \mbox{trace} \left( C^T {\cal H} C \right)
\end{equation}
and 
\begin{equation}
C^T C = I_m, 
\end{equation}
respectively. 
These equations are invariant under any unitary transformation of $C$. 

Although we focus on the standard eigenvalue problem of Eq.(\ref{BASICEQ}) in this work, 
the extension to the generalized eigenvalue problem 
(${\cal H} \evec=\lambda {\cal S} \evec$) is straightforward 
if ${\cal S}$ is a symmetric, positive definite matrix \cite{TRMIN}. 
This property allows us to use nonorthogonal basis functions \cite{FEM1,REVFEM,SIESTA,PAO} with ease. 
Moreover, the trace minimization method is equally valid even if 
${\cal H}$ depends on the eigenvectors of ${\cal H}$ itself, 
as explained in $\S$9.4.3.4 of Ref.\cite{TMPL}. 

In Fig.\ref{TMFIG}, we illustrate the numerical implementation of the trace minimization method 
in which $E(C)$ is minimized directly with respect to $C$ 
using the nonlinear conjugate gradient method \cite{CG0,PTAAJ,EDSM}. 
This procedure is often referred to as a {\it direct energy minimization} 
in electronic structure community \cite{DFTREV1,DFTREV2,DFTREV3}. 
Here $C, G, P \in \Rvec^{N \times m}$, and a line minimization is performed along $P_i$ 
to determine $\alpha_i$ \cite{RCP,KV}. 
Moreover, $\gamma_i$ is given by 
\begin{equation}
\gamma_i = 
\left\{
\begin{array}{ll}
0 & \qquad i=0 \\
\noalign{\vskip0.2cm} 
\displaystyle{\frac{\mbox{trace}((G_i-G_{i-1})^T G_i)}{\mbox{trace}(G_{i-1}^T G_{i-1})}}  & \qquad i=1,2,...
\end{array}
\right.
\end{equation}
following the Polak-Ribiere formula \cite{RCP}. 

When $\| G_i \|$, the Frobenius norm of $G_i$, is sufficiently small, 
$E_i$ will be equal to $E_{\rm G}$, and  
\begin{equation}
{\cal H} C = C H
\end{equation}
will hold. Diagonalization of $H$ $\in \Rvec^{m \times m}$ 
will give the explicit eigenvalues and eigenvectors of ${\cal H}$, if necessary. 
The number of iterations to reach convergence is estimated by \cite{ANNETT} 
\begin{equation}
\label{NITER}
N_{\rm iter} \propto \frac{1}{\sqrt{\epsilon_{\rm gap}}}. 
\end{equation}
Therefore, a naive implementation of the trace minimization method fails 
in the limit of a vanishing gap. In this case, more complex (and thus more costly) algorithms 
\cite{EDFT1,SATO} should be employed to avoid the slow convergence. 

\begin{figure}[t]
  \begin{center}
  \includegraphics[width=13cm]{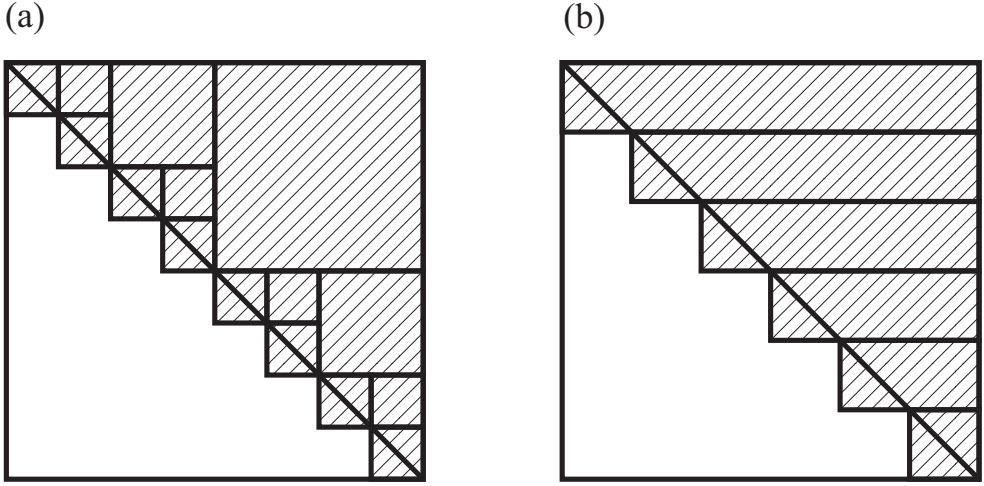}
  \end{center}
  \caption{(a) Recursive and (b) row-wise splitting of the upper triangular part of a symmetric matrix.}
  \label{DIVFIG}
\end{figure}

The procedure for calculating the energy and the gradient is illustrated in Fig.\ref{EGFIG}. 
Algorithm-1 is a naive implementation which is shown only for illustrative purposes. 
Algorithm-2 is a more practical implementation which is 
appropriate for incorporating the mixed precision arithmetic explained in $\S$\ref{MPSEC}. 
Since $H'$ is a symmetric matrix of dimension $m$, 
only the upper (or lower) triangular part needs to be calculated explicitly in step-(6). 
In particular, when $m$ is large, 
$H'$ should be divided into smaller blocks, as shown in Fig.\ref{DIVFIG}, 
with each block being calculated by a separate DGEMM (or SGEMM) call. 

Similarly, the orthonormalization procedure is shown in Fig.\ref{ONFIG}. 
Strictly speaking, explicit orthonormalization of $C$ is inconsistent 
with the conjugate gradient method, 
which is designed for unconstrained minimization problems. 
In our experience, however, only minor effects are seen when $ m \ll N$ is satisfied. 

Since all $O(m^2 N)$ operations introduced in this section 
are performed by level-3 BLAS routines, 
optimal performance is achieved on most platforms \cite{GBLAS}.

\begin{figure}[t]
Choose initial guess $C_0$ subject to $C_0^T C_0 = I_m$ \\
{\bf for} $ i = 0,1,2, ...$ \\
\hspace*{0.6cm} Calculate $E_i = E (C_i)$ and $G_i = -\nabla E (C_i)$ \\
\hspace*{0.6cm} check convergence; continue if necessary \\
\hspace*{0.6cm} $P_i = G_i + \gamma_i P_{i-1}$ \\
\hspace*{0.6cm} $C_{i+1} = C_i + \alpha_i P_i$ \\
\hspace*{0.6cm} Orthonormalize $(C_{i+1})$ \\
{\bf end}
\caption{Trace minimization method using the nonlinear conjugate gradient method.} 
\label{TMFIG}
\end{figure}

\begin{figure}[t]
Calculate $E (C)$ and $G = -\nabla E (C)$ \\
\begin{tabular}{lll}
 Algorithm-1: & & \\
(1) $X = {\cal H} C$ &  & $O(mN)$ \\
(2) $H = C^T X$ & DGEMM & $O(m^2 N)$ \\
(3) $E = {\mbox{trace}} (H)$ &  & $O(m)$ \\
(4) $G = -2 (X - C H)$ & DGEMM & $O(m^2 N)$ \\
 & & \\
 Algorithm-2: & & \\
(1) $X = {\cal H} C$ &  & $O(mN)$ \\
(2) $H_{\rm D} = \mbox{diag}(C^T X) $ &  & $O(m N)$ \\
(3) $E = {\mbox{trace}} (H_{\rm D})$ &  & $O(m)$ \\
(4) Quit if $G$ is unnecessary & & \\
(5) $X' = X - C H_{\rm D}$ &  & $O(mN)$ \\
(6) $H' = C^T X'$ & DGEMM  & $O(m^2 N)$ \\
(7) $G = -2 (X' - C H')$ & DGEMM & $O(m^2 N)$
\end{tabular}
\caption{Two (mathematically) equivalent procedures for calculating the energy and gradient. 
The second and third columns show the corresponding BLAS/LAPACK routines (if any) 
and their computational costs. 
$H, H', H_{\rm D}$ are symmetric matrices $\in \Rvec^{m \times m}$, and $X, X' \in \Rvec^{N \times m}$.}
\label{EGFIG}
\end{figure}

\begin{figure}[t]
Orthonormalize $(C_{\rm in})$: \\
\begin{tabular}{lll}
(1) $S = C^T_{\rm in} C_{\rm in}$ & DSYRK & $O(m^2 N)$ \\
(2) Calculate $L$, where $S = L L^T$ & DPOTRF & $O(m^3)$ \\
(3) Calculate $L^{-1}$  & DTRTRI & $O(m^3)$ \\
(4) $C_{\rm out} = C_{\rm in} L^{-1}$ & DTRMM & $O(m^2 N)$
\end{tabular}
\caption{
Orthonormalization procedure based on the Cholesky factorization. 
Same notation as in Fig.\ref{EGFIG}. 
$S$ is a symmetric matrix of dimension $m$, and 
$L$ is a lower triangular matrix of dimension $m$ (so is $L^{-1}$). 
On exit, $C_{\rm in}$ is overwritten by $C_{\rm out}$.}
\label{ONFIG}
\end{figure}

\subsection{Mixed precision arithmetic}
\label{MPSEC}
Here we present several ideas for improving the performance of the algorithm 
introduced in $\S$\ref{TMMSEC} by taking advantage of inexpensive SP operations. 
The first approach is to perform all floating-point operations in DP ({\bf full DP}), 
which will serve as a reference in the following. 
On the other hand, it is also possible to replace all DP operations by SP ({\bf full SP}), 
which is expected to achieve the largest gain 
in terms of computational cost and memory usage. 
Unfortunately, as will be shown below, this approach does not provide sufficient accuracy, 
and thus is of limited use in real applications. 

A more practical approach is to start from full DP, 
and incorporate SP operations progressively. 
Mixed precision variant-1 ({\bf MP1}) is a conservative approach 
which aims at achieving a reasonable gain, while keeping full DP accuracy. \\
(i) The change of $C_i$, $\|C_{i+1}-C_i\|$, in Fig.\ref{TMFIG}, will become much smaller than $\| C_i \|$ 
as convergence is approached. 
Therefore, $G_i$ and $P_i$ can be stored in memory in SP format without sacrificing accuracy. 
Here, the elements of $G_i$ are first calculated in DP 
to minimize round-off errors in Fig.\ref{EGFIG}, followed by conversion to SP. 
Conversion between SP and DP is performed with the Fortran 
intrinsic functions {\sc Sngl} and {\sc Dble}. 
While this change alone leads to only a minor performance gain, 
a large gain can be made when used in conjunction 
with advanced preconditioners, as will be discussed in $\S$\ref{DISSEC}. 
Furthermore, a significant reduction of memory usage is expected. \\
(ii) Similarly, the change of $C_i$ during the orthonormalization procedure 
shown in Fig.\ref{ONFIG} decreases from iteration to iteration. 
Therefore, after calculating $L^{-1}$, we introduce a DP matrix 
\begin{equation}
L_{\rm D} = \mbox{diag} (L^{-1}), 
\end{equation}
and an SP matrix 
\begin{equation}
L' = L^{-1}-L_{\rm D},
\end{equation}
where $L_{\rm D} \rightarrow I_m$ and $L' \rightarrow 0$ as convergence is approached. 
Then, the last step can be rewritten as 
\begin{equation}
C_{\rm out} = C_{\rm in} L_{\rm D} + C_{\rm in} L',
\end{equation}
where the first and the second terms are calculated in DP and SP, 
respectively. Obviously, the former cost is negligible, 
while the latter can be performed with an STRMM call instead of DTRMM. 
The final result, $C_{\rm out}$, is stored in DP format. 

In addition to the changes noted above, further acceleration is achieved 
in the mixed precision variant-2 ({\bf MP2}) 
by reducing the cost of evaluating the gradient in Fig.\ref{EGFIG}, Algorithm-2, as follows. 
Since the computational cost of this procedure is dominated by steps-(1), (6), and (7), 
the two DGEMM calls in steps-(6) and (7) are replaced by SGEMM. 
The rest of the operations in this procedure, including step-(1), are performed in DP, 
which guarantees DP accuracy of the energy. 
Step-(5), corresponding to self-orthogonalization of the gradient, 
is also performed in DP, which allows us to reduce 
the round-off errors arising from the use of SGEMM in steps-(6) and (7) significantly. 
It is also preferable to set diag($H'$)=0 explicitly 
after step-(6) to minimize the errors. 
Nevertheless, MP2 has the potential risk of obtaining inaccurate gradient when close to convergence. 

When MP2 is used, all $O(m^2 N)$ operations except the DSYRK call in the orthonormalization procedure 
are performed in SP. 
The performance and accuracy of these algorithms are compared in the next section.

\section{Numerical results}

\begin{figure}[t]
  \begin{center}
  \includegraphics[width=9cm]{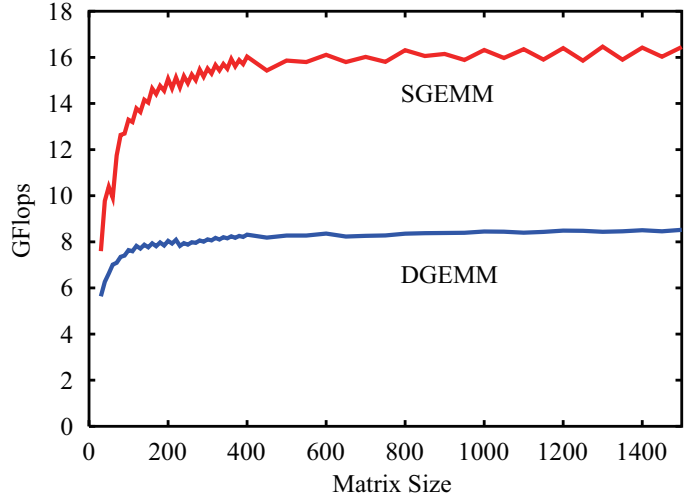}
  \end{center}
  \caption{Performance of matrix-matrix multiplications in SP (SGEMM) and DP (DGEMM).}
  \label{MMFIG}
\end{figure}

All calculations shown here were performed on a single core of 
the 2.3 GHz AMD Opteron 6176 SE processor with CentOS 5.5 operating system, 
gfortran 4.1.2 compiler, and GotoBLAS2 numerical library \cite{GBLAS}. 

We first show the performance of matrix-matrix multiplications in Fig.\ref{MMFIG}. 
The SP routine (SGEMM) is found to be 1.85-1.95 times faster than 
the corresponding DP routine (DGEMM) when the matrix size is larger than 200. 

\begin{figure}[t]
  \begin{center}
  \includegraphics[width=9cm]{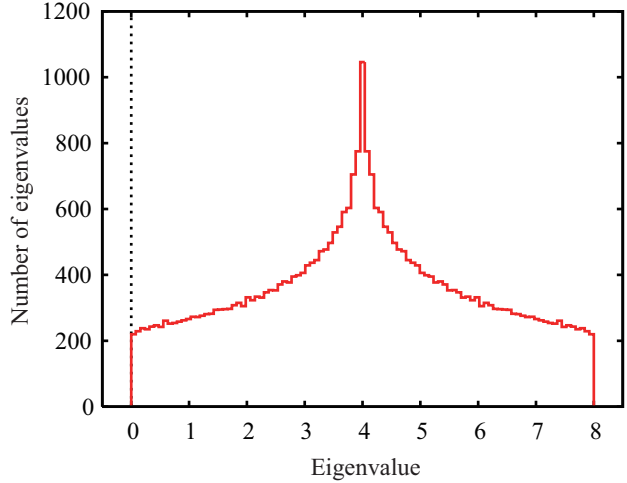}
  \end{center}
  \caption{The eigenvalue distribution of ${\cal H}_N$ (Eq.(\ref{LAPLACE})) for $N=192^2$.}
  \label{DOSFIG}
\end{figure}

Then, we compare the performance of full DP, MP1, MP2, and full SP 
for calculating the sum of the $m$ lowest eigenvalues of 
large sparse matrices corresponding to the two-dimensional 
discrete Laplacian under Dirichlet boundary conditions, 
\begin{equation}
\label{LAPLACE}
{\cal H}_N = \left(
\begin{array}{ccccc}
 D_n & -I_n &  &  & \bigzerou \\
-I_n &  D_n & -I_n &  &  \\
    & \ldots  & \ldots & \ldots &  \\
    &  & -I_n & D_n & -I_n \\
\bigzerol &  &   & -I_n & D_n \\
\end{array}
\right) \in \Rvec^{N \times N},
\end{equation}
\begin{equation}
D_n = \left(
\begin{array}{ccccc}
 4 & -1 &    &  & \bigzerou \\
-1 &  4 & -1 &  &  \\
   &  \ldots & \ldots & \ldots &  \\
   &  & -1 & 4 & -1 \\
\bigzerol &  &    & -1 & 4 \\
\end{array}
\right) \in \Rvec^{n \times n}. 
\end{equation}
The dimension $N$ is given by $N=n^2$, where $n$ denotes 
the grid size in each direction. 
The eigenvalues of ${\cal H}_N$ are given explicitly by 
\begin{equation}
\lambda_{p,q} = 4 \left( \sin^2{\left(\frac{p \pi}{2(n+1)}\right)} + 
\sin^2{\left(\frac{q \pi}{2(n+1)}\right)}\right), \quad \quad p,q=1, 2, ...,n. 
\end{equation}
After sorting $\left\{ \lambda_{p,q} \right\}$ in ascending order, 
we can calculate the exact value of $E_{\rm G}$ for any value of $m$. 
The above Hamiltonian represents free electrons confined in a square box, 
which is essentially a gapless system, as shown in Fig.\ref{DOSFIG}. 
Therefore, this problem is a stringent test for the trace minimization method 
which requires the presence of an energy gap.

\begin{table}[t]
\caption{Performance of the four algorithms 
for iterative diagonalization of ${\cal H}_N$. 
$N_{\rm iter}$ denotes the number of iterations required to achieve $R < 10^{-12}$ in the full DP run. 
All measurements are given in units of seconds per iteration. 
The numbers in parentheses indicate the percentage relative to full DP.} 
\label{RESTBL}
\begin{center}
\begin{tabular}{cccccccccc}
\hline
\hline
$N$ & $m$ & $E_{\rm G}$ & $\epsilon_{\rm gap}$ & $N_{\rm iter}$ & Full DP & MP1 & MP2 & Full SP \\
\hline
 96$^2$ & 220 & 35.25 & 6.2$\times$10$^{-3}$ & 270 & 0.631 & 0.591(94\%) & 0.491(78\%) & 0.419(66\%) \\
192$^2$ & 220 & 8.991 & 1.9$\times$10$^{-3}$ & 630 & 2.520 & 2.357(94\%) & 1.968(78\%) & 1.680(67\%) \\
192$^2$ & 534 & 50.90 & 2.5$\times$10$^{-3}$ & 560 & 12.04 & 11.05(92\%) & 8.772(73\%) & 7.958(66\%) \\
192$^2$ & 1064& 196.8 & 3.4$\times$10$^{-3}$ & 460 & 45.38 & 41.01(90\%) & 30.52(67\%) & 28.04(62\%) \\
192$^2$ & 1519& 395.6 & 2.7$\times$10$^{-3}$ & 422 & 89.84 & 79.16(88\%) & 60.53(67\%) & 57.23(64\%) \\
\hline
\hline
\end{tabular}
\end{center}
\end{table}

\begin{figure}[t]
  \begin{center}
  \includegraphics[width=15cm]{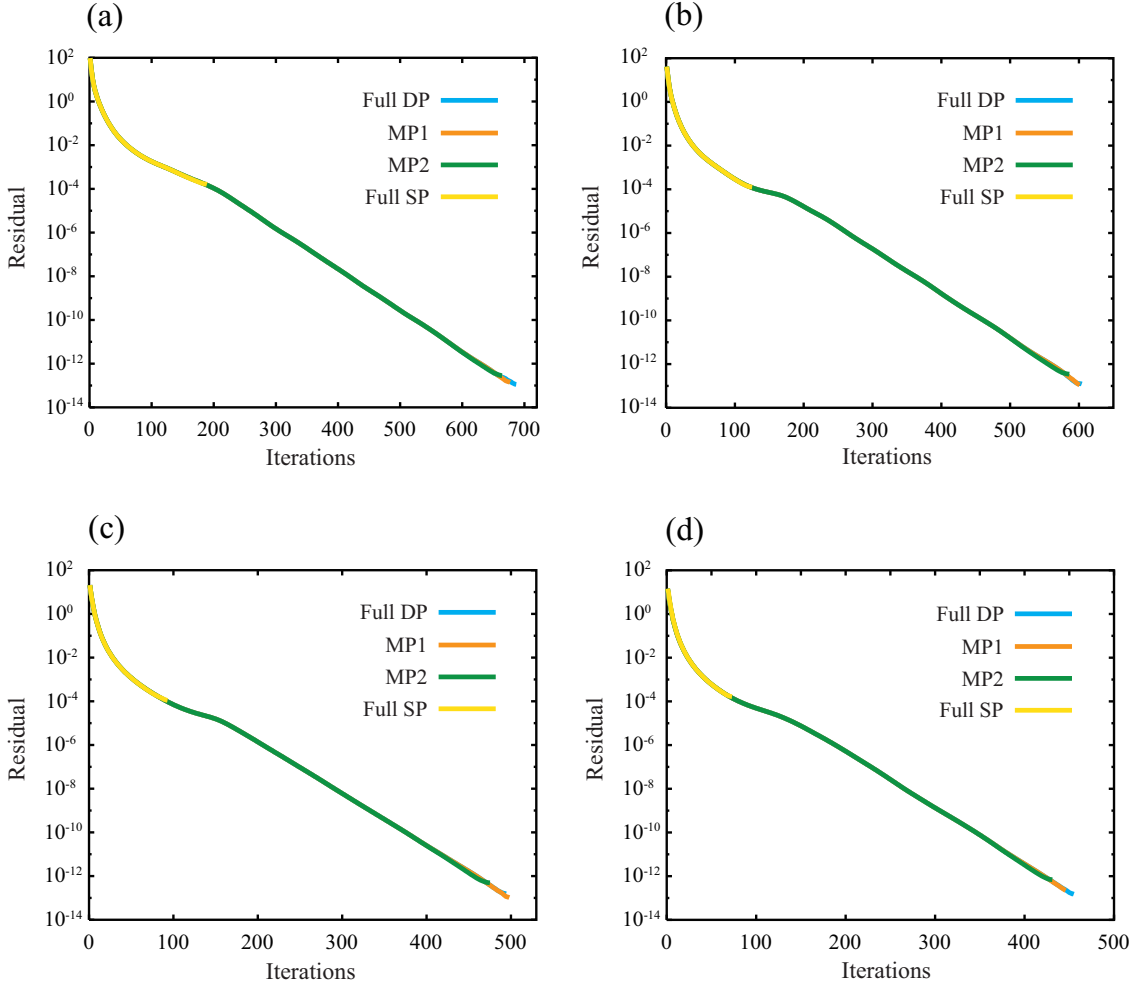}
  \end{center}
  \caption{Convergence of the residual for $N=192^2$: (a) $m=220$, (b) $m=534$, 
	(c) $m=1064$, and (d) $m=1519$.}
  \label{CONVFIG}
\end{figure}

In Table \ref{RESTBL}, we show the results of iterative diagonalization 
for several pairs of ($N,m$), following the algorithm presented in $\S$\ref{TMMSEC}. 
Here, the values of $m$ were chosen to satisfy $\epsilon_{\rm gap} > 0$. 
For simplicity, the symmetry of $H'$ was not exploited, and 
the initial guess $C_0$ was generated from random numbers, followed by orthonormalization. 
In Fig.\ref{CONVFIG}, we show the convergence of the residual at iteration $i$, defined by 
\begin{equation} 
R_i = \frac{E_i - E_{\rm G}}{E_{\rm G}}. 
\end{equation} 
These results suggest that the performance of the four algorithms satisfies 
$$ \mbox{Full SP} \, > \, \mbox{MP2} \, > \, \mbox{MP1} \, > \, \mbox{Full DP}, $$ 
where the differences tend to increase with $m$, but not with $N$. 
In particular, full SP is 30 - 40 \% faster than full DP, 
which is consistent with the results for matrix multiplications. 
However, the accuracy of the results obtained from full SP 
is insufficient for most applications. 
Therefore, full SP should be used only for generating the initial guess \cite{KOTA}. 

In contrast, MP1 retains full DP accuracy for all values of $(N,m)$ shown in Table \ref{RESTBL}, 
while showing only a modest gain of $\approx$ 10 \%. 

MP2 is found to achieve a gain of over 30 \% for large $m$, while keeping near DP accuracy. 
However, the accuracy of the converged solution deteriorates slowly with $m$, 
because the use of SP is not always appropriate for evaluating the gradient. 
We have found that if subspace rotation is performed occasionally to 
diagonalize $H$, this problem can be overcome, thus leading to full DP accuracy. 
Alternatively, one can simply switch from MP2 to MP1 (or full DP) algorithm 
when the residual is below a given tolerance. 
The latter approach is preferable in terms of the construction of 
conjugate directions \cite{RCP}, as well as the extrapolation of the initial guess \cite{APJ}.

\section{Discussion}
\label{DISSEC}

In $\S$\ref{TMMSEC}, we illustrated the implementation of the trace minimization method using 
the basic conjugate gradient method. 
In this section, we show how to improve the convergence rate of the conjugate gradient method 
by a linear transformation of the gradient 
\begin{equation}
\label{PCEQ}
G' = M G.
\end{equation}
Here, $M \in \Rvec^{N \times N}$ is a symmetric, positive definite matrix 
called a preconditioner, which should be a reasonable 
approximation to the inverse Hessian (See $\S$11 of Ref.\cite{TMPL}). 
Preconditioning allows us to reduce the number of iterations to reach convergence 
at the expense of increased cost per iteration. 
In electronic structure calculations, 
this generally leads to an increase in $O(m N)$ cost and a decrease in $O(m^2 N)$ cost. 
Therefore, the choice of the preconditioner becomes more important for larger applications. 
In plane-wave-based electronic structure calculations, 
it is a common practice to use a diagonal matrix for $M$, 
which leads to a considerable reduction of the number of iterations \cite{PTAAJ} 
at a negligible cost. 

However, more elaborate preconditioners have been developed beyond the diagonal approximation 
for plane waves \cite{SWMR}, atomic orbitals \cite{WVHN}, and 
real space basis sets \cite{MG1,MG2,FA}. 
These preconditioners significantly improve the convergence rate, 
while the computational cost of evaluating the right-hand side of Eq.(\ref{PCEQ}) 
will become non-negligible. 
Since SP is generally sufficient for representing $G$, as already mentioned in $\S$\ref{MPSEC}, 
the same will hold for $M$, unless $M$ is an ill-conditioned matrix. 
Therefore, the matrix multiplication of Eq.(\ref{PCEQ}) can also be performed in SP, 
thus minimizing the overhead of preconditioning. 
The idea behind this approach is similar in spirit to 
the solution of linear equations discussed in Ref.\cite{LIN2}. 
These advanced preconditioners will be particularly beneficial 
when highly accurate methods \cite{GSHV} are used for the electronic structure calculations. 

Although we have focused on the conjugate gradient method, 
the basic idea presented in this work should be equally valid 
for other iterative algorithms \cite{DAVID,DIIS,JDM,OZ2,BOMI}. 
In our electronic structure code {\sc Femteck} \cite{FEM1,FEM2}, 
the trace minimization method is used in conjunction with 
the limited memory BFGS (Broyden-Fletcher-Goldfarb-Shanno) method \cite{HEPO,LINO,QNFEM,LMRHM}, 
which requires practically no line minimization. 
When the MP2 algorithm is used, together with a multigrid approximation to the inverse Hessian \cite{FA}, 
the computational time for the ground-state calculation is 
reduced by a factor of about two in large-scale applications \cite{SPES}, 
compared with the full DP calculation using a diagonal approximation to the Hessian \cite{QNFEM,LMRHM}. 
If we start from a reasonable initial guess \cite{APJ}, 
the ground-state energy is obtained in 10-15 iterations 
in nonmetallic materials \cite{NAFION,SPES}.

\section{Conclusion}
We have shown in this work that iterative solution of the eigenvalue problem 
can be accelerated significantly by taking advantage of mixed precision arithmetic 
without relying on external devices. 
Even further improvement can be made by using problem-specific preconditioners 
which take into account nondiagonal elements. 
These methods will become more important with increasing problem size. 

For most applications, MP2 is a good compromise between accuracy and computational cost. 
When highly accurate eigenvalues/vectors are desired, 
we recommend to switch from MP2 to MP1 (or full DP) 
at some point before the convergence slows down. 

While the current implementation is designed for ground-state electronic structure calculations, 
it would also be interesting to investigate the use of mixed precision arithmetic 
in other problems such as the density-functional perturbation theory \cite{MBPT1,MBPT2,MBPT3}.

\section*{Acknowledgements}
This work has been supported in part by a KAKENHI grant (22104001) from 
the Ministry of Education, Culture, Sports, Science and Technology, 
and a grant from the Ministry of Economy, Trade, and Industry, Japan.

\end{document}